\documentclass{aastex}
\usepackage{emulateapj5}

\newcommand{\etal}{{et al.}\hspace{1mm}}

\begin{document}


\title{Explaining the entropy excess in clusters and groups of
galaxies without additional heating}

\author{Greg L. Bryan\altaffilmark{1}}
\affil{Department of Physics, Massachusetts Institute of Technology,
Cambridge, MA 02139}
\email{gbryan@mit.edu}


\altaffiltext{1}{Hubble Fellow}


\begin{abstract}

The X-ray luminosity and temperature of clusters and groups of
galaxies do not scale in a self-similar manner.  This has often been
interpreted as a sign that the intracluster medium has been
substantially heated by non-gravitational sources.  In this paper, we
propose a simple model which, instead, uses the properties of galaxy
formation to explain the observations.  Drawing on available
observations, we show that there is evidence that the efficiency of
galaxy formation was higher in groups than in clusters.  If confirmed,
this would deplete the low-entropy gas in groups, increase their
central entropy and decrease their X-ray luminosity.  A simple,
empirical, hydrostatic model appears to match both the
luminosity-temperature relation of clusters and properties of their
internal structure as well.

\end{abstract}

\keywords{cosmology: theory, intergalactic medium}



\section{Introduction}

Clusters and groups of galaxies are composed of galaxies, hot X-ray
emitting gas, and a gravitationally dominant dark halo.  Although this
basic picture is well-understood, there remain a number of puzzles
that prevent clusters from being fully used as standard candles.  For
example, simple scaling relations (and detailed simulations) predict
that the bolometric X-ray luminosity should scale with the temperature
of the gas according to $L_X \sim T^2$, while observations indicate
$L_X \sim T^3$ \citep{kai91, evr91,nav95,bry98}. Early on it was
suggested that groups might have a lower baryon fraction because of
more efficient star formation \citep[e.g.][]{dav92,tho92,bow97}, but
later thinking has focussed on the idea of additional heating of the
gas, often assumed to be from supernovae or active galactic nuclei
\citep[e.g.][]{kai91,bow00}.  This non-gravitational heating would
decrease the central density and, because the X-ray emissivity is
proportional to the density squared, reduce the luminosity.  Because
of the lower pressures in smaller clusters and groups, this would
preferentially affect them, steepening the $L_X-T$ relation
\citep{cav97}.  This viewpoint was strengthened by the discovery of an
apparent entropy ``floor'' in the centers of groups and clusters
\citep{pon99}.  This is consistent with the idea of a source of heat
which raises the entropy of the gas to a fixed, minimum level; smaller
clusters have lower entropies and so are more affected than large
clusters.  This model has been developed in some detail in a number of
papers \citep{bal99,loe00,wu99,cav99,val99} and appears to be capable
of naturally explaining the observations.

The amount of heating required is substantial.  Although estimates
vary, it seems likely that about 1 keV per particle is needed
\citep{llo00}, an amount which may be challenging to explain from
supernova heating alone \citep{kra00}.  Another difficulty is that
observations of the Ly$\alpha$ forest indicate a much lower
temperature for the majority of the intergalactic medium at $z \sim
2-3$ \citep{bry00,sch99}, a condition which may extend to
even lower redshift \citep{ric00}.  Although hardly conclusive, these
concerns may be pointing toward another explanation for the
observations.

This paper argues that cooling and the resultant galaxy formation are
sufficient, by themselves, to explain all of these observations and
that substantial heating is not required.  We draw mostly on two
simple ideas: (1) small clusters and groups have converted more of
their baryonic gas into galaxies than have large clusters and (2) the
gas which goes to form the galaxies is preferentially lower in
entropy, thus raising the mean entropy of the gas which remains.  This
means that not only do small clusters have a smaller gas fraction
($f_{gas}$), the gas which is there has a higher entropy --- and lower
density --- than it would in simple self-similar scaling models.
Because of the density-squared nature of the X-ray emission, this
substantially diminishes the luminosity of groups and small clusters,
resulting in a steeper $L_X-T$ relation, as observed.  The effective
entropy increase is most noticeable in the center of the cluster,
which is just where the entropy floor is observed.  In what
follows, we explore a simple model to investigate if this hypothesis
can match the observations quantitatively.  We will also show that
there is some empirical support for the first assumption.



\section{The Model}
\label{sec:model}

The model described here is built on the assumption that galaxy
formation is not uniformly efficient in all environments.  Since
theoretical arguments can and have been made both ways, we attempt to
address this point with observations drawn from the literature.  We
searched the literature and found three studies that computed stellar,
gas and total masses within the same radius (many more computed a
subset of these three quantities but we only used those that computed
all of the quantities to insure self-consistency).  Mulchaey \etal
(1996) used their own ROSAT and optical observations combined with
other results from the literature to compile a list of 16 groups with
masses computed out to $R_X$, the maximum radius at which X-ray
emission could be observed.  Hwang \etal (1999) used ASCA observations
of five intermediate mass systems and also computed masses out to
$R_X$.  Finally, Cirimele \etal (1997) studied 12 Abell clusters with
ROSAT, and tabulate masses computed out to 1.5 Mpc (which is close to
$R_X$ for their clusters).  All studies used similar, although not
identical, stellar mass to light ratios (corrected for morphological
variations) and all used the same cosmological parameters.

In Figure~\ref{fig:mratio}, we plot the relative stellar and gas
masses from these three studies.  This shows that
the hot gas component dominates over galaxies in the most massive
clusters of galaxies.  For smaller systems, the scatter increases
significantly; however, there is a trend towards an increasing stellar
contribution and decreasing gas contribution for lower mass clusters
and groups.  To make this clearer, we divide the sample into four
equal-sized groups, ordered by temperature, and plot the median for
each group.  To check the statistical strength of the trend, we
divided the sample into two groups (those with temperatures below and
above 2 keV), and separately computed the median galaxy mass ratios.
Then, using median statistics (e.g. Gott \etal 2000), we find the
probability that the median of the high temperature group is smaller
than that of the low temperature group to be 0.986.  The gas mass
ratio trend was even more significant.

\vspace{\baselineskip}
\epsfxsize=3.5in
\centerline{\epsfbox{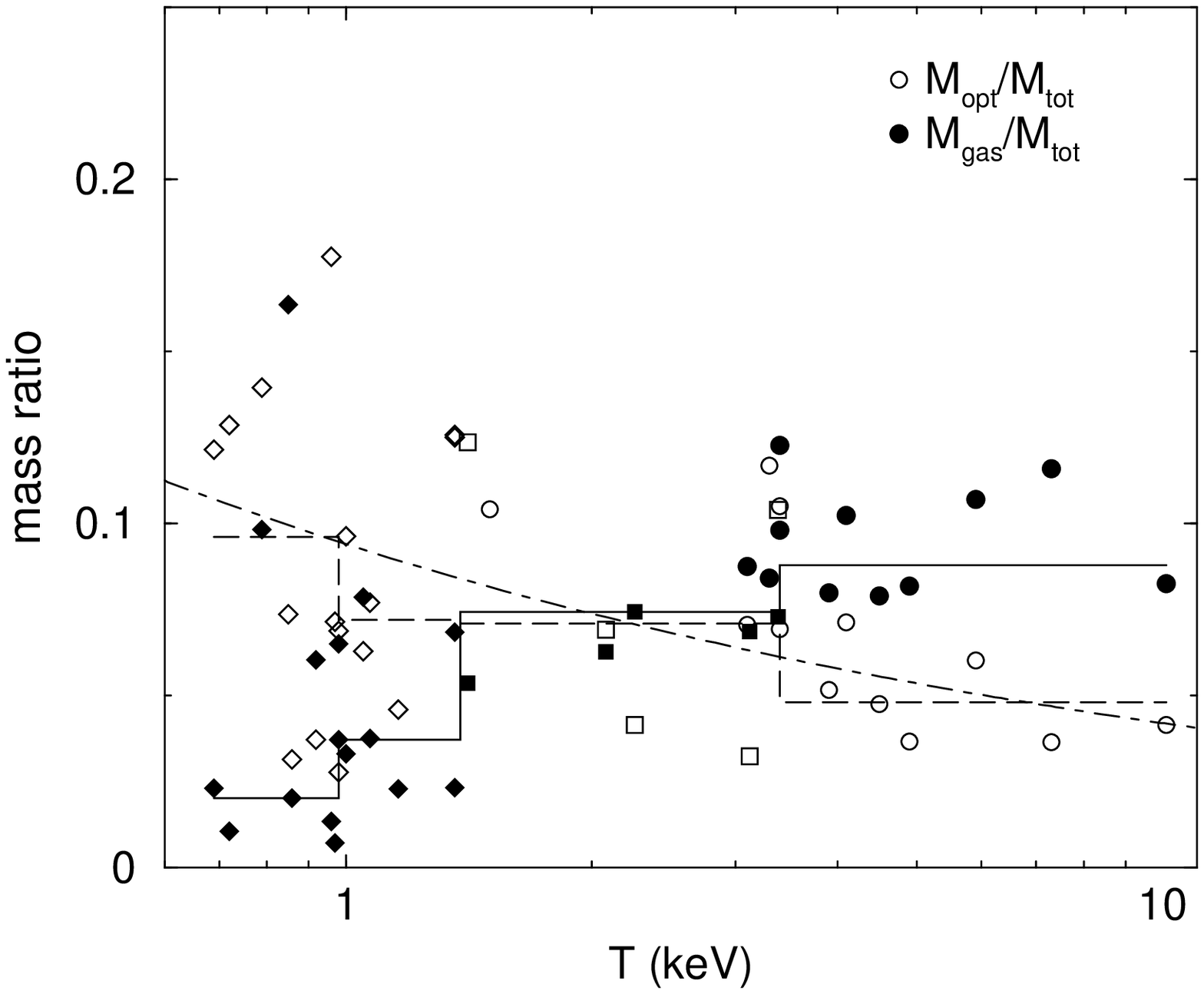}}
\figcaption{The ratio of mass in hot gas (solid symbols) and in stellar
systems (open symbols) to total gravitating mass as a function of
temperature.  The circles are from Cirimele, Nesci \& Tr\'evese
(1997), the squares from Hwang \etal (1999)
and the diamonds from Mulchaey \etal (1996).  Solid and dashed lines are
median values of the binned distribution for the gas ratio and stellar
ratio, respectively.  The dot-dashed line is the model discussed in
the text.  All observations have been adjusted to $h=0.65$.
\label{fig:mratio}}
\vspace{\baselineskip}

The most straightforward explanation of the trend in
Figure~\ref{fig:mratio} is that the efficiency of galaxy formation
varies from groups to clusters.  This is the hypothesis that we will
examine in this paper, but there are certainly other explanations.
For example, it is possible that the relative mix of gas and stars
changes outside of the measured region (i.e. $R > R_X$), which is
generally a smaller fraction of the virial radius for groups than for
clusters.
Moreover, it is difficult to be conclusive for a heterogeneous sample
of this sort, since the groups and clusters were examined by different
authors using slightly different methods.  However, all of the studies
did use the same basic methodology, and adopted similar parameters.
Also, the trend itself does not depend on correctly determining the
total mass, since the ratio $M_{gas}/M_{opt}$ --- which is independent
of total mass --- also decreases with temperature.

There are a number of other pieces of evidence which support this
basic conclusion.  For example, Arnaud \& Evrard (1999) and Mohr,
Matheison \& Evrard (1999) find a trend of decreasing hot gas fraction
with decreasing temperature for their samples.  This pattern has
sometimes been taken to imply that gas has been ejected from smaller
clusters and groups, despite the large amount of energy required to do
this.  However, it seems equally possible that this gas is in the form
of stars.  This would also be consistent with a higher mass-to-light
ratio for large clusters than for groups
\citep{gir00,ada98,hra00,ram97}, although see \citet{dav95}.  Weak
lensing of groups should provide useful constraints; preliminary
results indicate the the mass-to-light ratios of groups are somewhat
lower than clusters \citep{hoe00}.  Finally, constraints from galaxy
clustering indicates the number of galaxies in a halo must grow more
slowly then the mass of the halo \citep{sco00,sel00}, consistent with the
trend presented here.  Despite this circumstantial evidence, we cannot
prove that the efficiency of galaxy formation depends on environment;
all we can do is show that the available data are consistent with the
trend shown in Figure~1.  In the rest of the paper, we will assume
this to be true and examine the consequences that follow.

In Figure~\ref{fig:mratio}, we show the relation, 
\begin{equation}
f_{star} = 0.042(T/10\hbox{keV})^{-0.35}
\label{eq:fstar}
\end{equation}
which we will take to be the stellar mass fraction in this paper, and
is the result of minimizing the absolute deviation of the mass ratio.
The gas fraction is simply $f_{gas} = f_{baryon} - f_{star}$, where
$f_{baryon} = 0.16$ is compatible with the cosmological model we have
chosen.  Specifically, this is a flat model with
($\Omega_0$,$\Omega_b$,$h$) = (0.35,0.056,0.65), where $\Omega_0$ is
the ratio of the mass density to the critical density and $h$ is the
Hubble constant in units of 100 km/s/Mpc.  The results are most
sensitive to the value of the Hubble constant since the ratio of
stellar to gas mass varies as $h^{-3/2}$; the other parameters play
almost no role.

In order to build a concrete model for the structure of a group or
cluster of galaxies we assume that: (1) the clusters are spherically
symmetric and in hydrostatic equilibrium; (2) the hot gas and stellar
fractions are as given above; (3) the gas which is converted into
stars comes from the lowest entropy gas in the cluster and all other
fluid elements lie on the same adiabat they would have without cooling
or star formation.  From experience gained with numerical simulations,
we know that while clusters are not in exact hydrostatic equilibrium
this assumption is a reasonable approximation.  The second assumption
has some empirical basis, as previously discussed.  We will return to
a discussion of the last assumption.

To create a cluster of a given mass $M$, we assume the dark matter density 
is described by \citep{nav96}:
\begin{equation}
\frac{\rho(x)}{\rho_0} = \frac{200}{3}\frac{c^2}{\ln(1+c)- c/(1+c)}
\frac{1}{x(1+cx)^2}
\label{eq:NFW}
\end{equation}
where $\rho_0 = 3H^2/8 \pi G$ is the critical density, $x = r/r_{200}$
and $c$ is a concentration parameter that depends weakly on mass in
the range of interest.  We take $c=8.5(Mh/10^{15}
M_{\odot})^{-0.086}$. The radius $r_{200}$ is defined by $M = 800
\rho_0 \pi r_{200}^3/3$.  

\vspace{\baselineskip}
\epsfxsize=3.5in
\centerline{\epsfbox{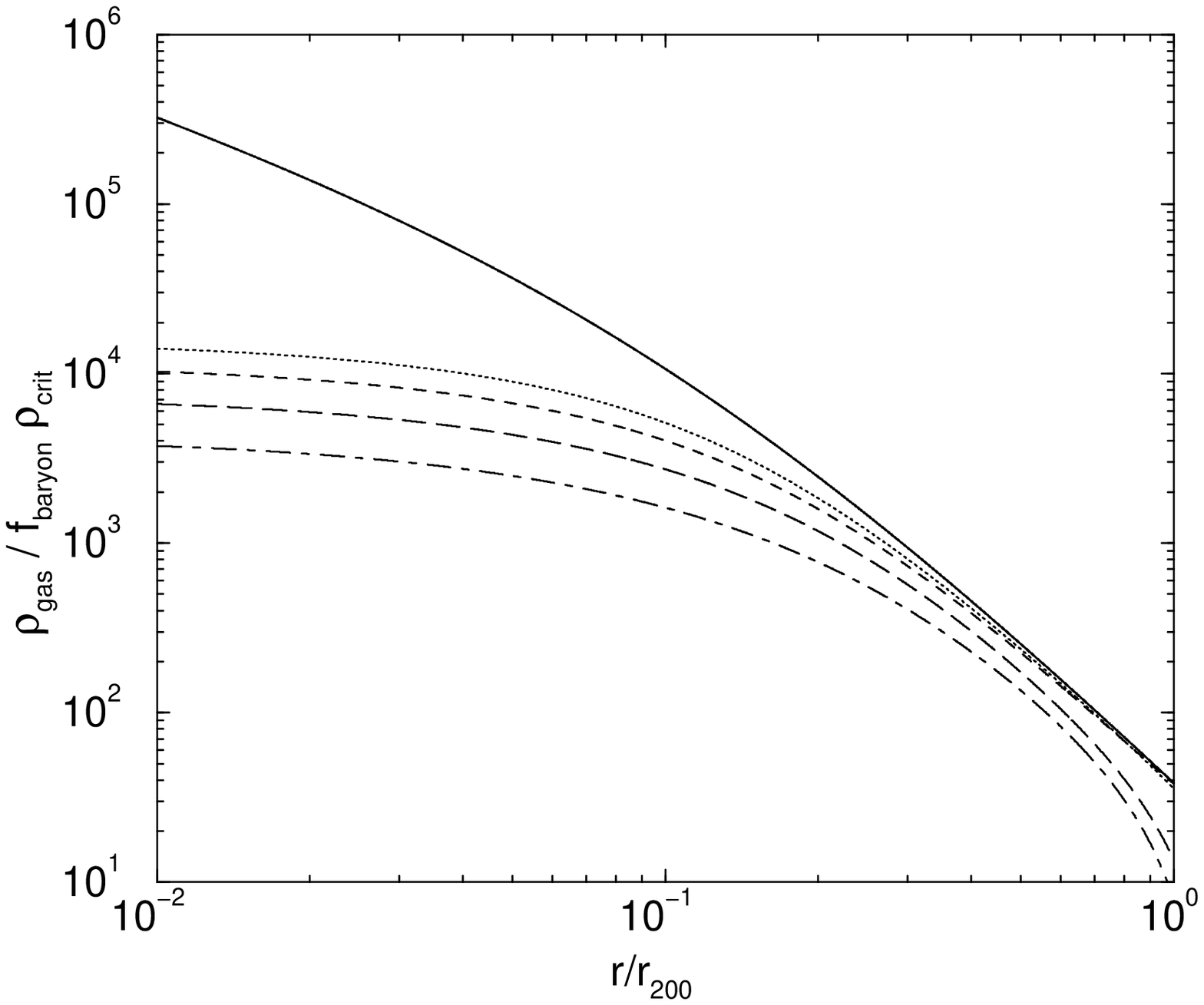}}
\figcaption{The gas profiles of clusters after galaxy formation for
temperature ranging from 10 keV (dotted) to 1 keV (dot-dashed).  The top
(solid) curve shows the initial profile.  Note that the decrease
with cluster temperature is entirely due to the increased efficiency
of galaxy formation -- without this all curves would look like the
initial profile.
\label{fig:profiles}}
\vspace{\baselineskip}

The gas distribution without galaxy formation is assumed to have the
same distribution as eq.(\ref{eq:NFW}).  The temperature profile is
determined by solving the equation of hydrostatic equilibrium for a
spherical profile: $dP/dr = - \rho_b GM(r)/r^2$.
We assume that the gas does not contribute to the gravitational
potential and adopt a pressure-free external boundary condition
(i.e. $P=0$ at $x=1$).  The result matches the density and temperature
distribution in numerical simulations over the vast majority of the
cluster \citep[e.g.][]{fre00}.  It fails in the very center (where the
numerical models are the least certain) but this represents a
small fraction of the mass ($< 1$\%) which will end up
being converted to galaxies anyway.
An isothermal temperature profile produces results which are broadly
similar;
however, this density distribution does not well describe the
simulation results, particularly at large radius ($x > 0.3$) where
much of the mass resides.

Once the no-cooling cluster has been constructed, we can then compute
the structure of the cluster including galaxy formation.  The equation
of hydrostatic equilibrium sets the
pressure distribution, but we need one more constraint to uniquely fix
the density and temperature profiles.  This comes from the entropy
($S=\ln T / \rho_b^{2/3}$) distribution of the gas, which
is a monotonically increasing function of radius.  Since by our
earlier assumption, the gas which cools into galaxies comes from the
lowest entropy part of the distribution, galaxy formation is
equivalent to removing from the center an amount of gas equal to
$f_{star} M$.  The remaining gas is then distributed over the whole
cluster, under the assumption that it does not cool at all.  The known
entropy distribution of this gas combined with the equation of
hydrostatic equilibrium is sufficient to specify the gas and
temperature profiles uniquely.  More precisely, we guess a central
pressure and then integrate the equation of hydrostatic equilibrium
outwards in radius, or equivalently, enclosed baryonic mass $M_b(r)$.
At each point, the density and temperature is computed from the
pressure and entropy $S(M_b)$, with the entropy coming from the
no-cooling case.  We iterate this procedure with a new central
pressure each time until a profile is produced which conserves mass
(i.e. $M_b(r_{200}) = f_{baryon}M$). Although this may superficially
appear to be a cooling-flow model, it is not.  The gas that cools into
galaxies does so at high redshift, before the cluster even forms.  The
gas which goes into the galaxies is at the center of the smaller halos
which merge to form a cluster.  If that gas hadn't cooled to form
galaxies, it would have ended up in the center of the cluster, since
low entropy gas will sink to the center.

Figure~\ref{fig:profiles} shows the resulting density profiles for a
range of cluster temperatures.  The temperature profiles remain mostly
flat, although they increase slowly towards the center and have a
somewhat larger mean compared to their non-cooling equivalents.

\vspace{\baselineskip}
\epsfxsize=3.5in
\centerline{\epsfbox{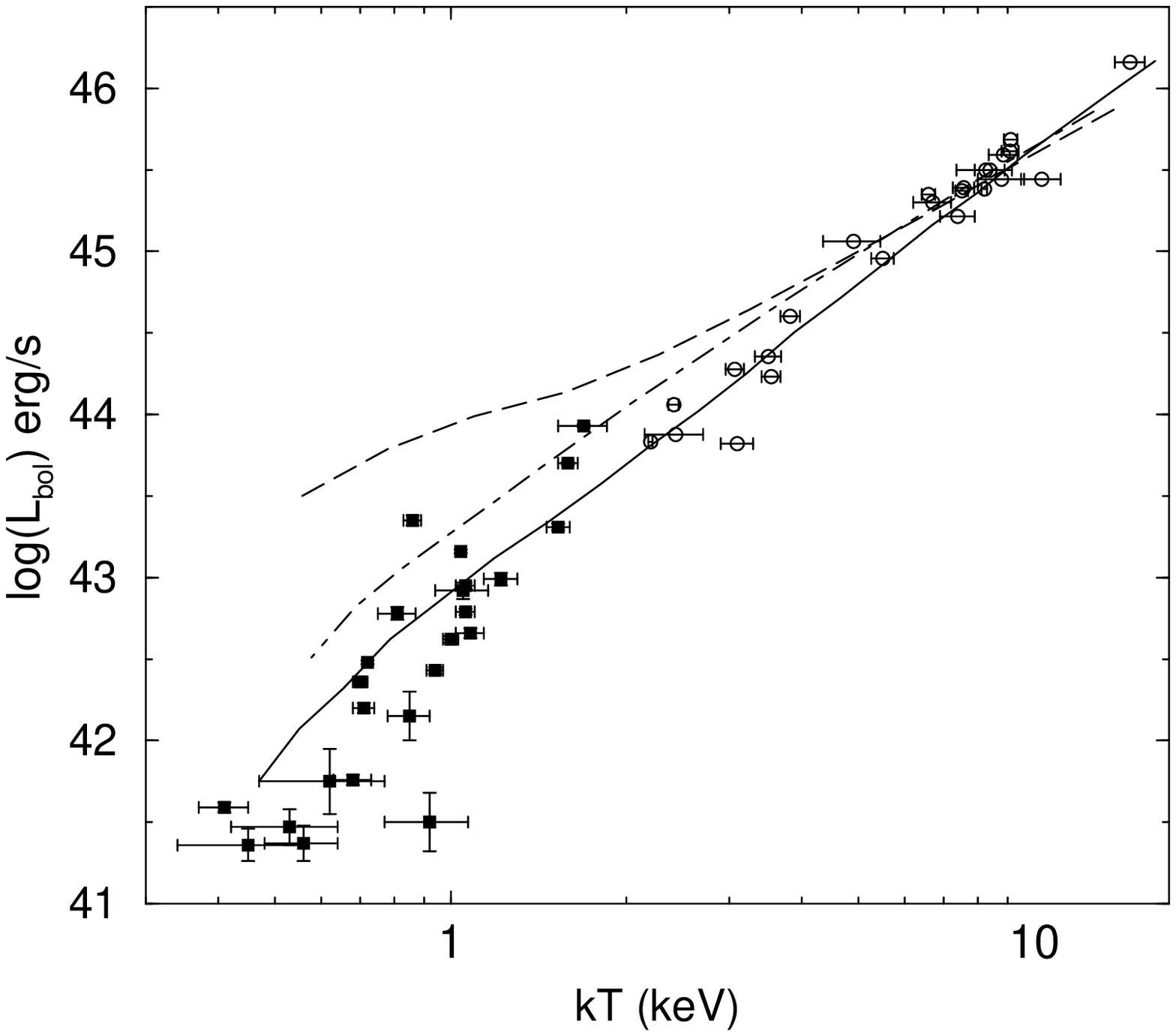}}
\figcaption{The data points show the observed bolometric X-ray
luminosity temperature relation, with open points from Arnaud \&
Evrard (1998) and filled points from Helsdon \& Ponman (2000).
The solid line is the $L_X-T$ relation for the cluster model
discussed in the text, while the dashed line is for a self-similar
model.  The dot-dashed curve shows the result of adopting the
variable $f_{gas}$ fraction, but using a self-similar profile for
the gas.
\label{fig:lum_temp}}
\vspace{\baselineskip}


\section{Comparison to observations}
\label{sec:observations}

The first test is to see if the resulting clusters can match the
luminosity-temperature relation.  For each cluster the X-ray
luminosity and luminosity-weighted temperature is computed with a
Raymond-Smith code (1977; 1992 version) assuming a constant
metallicity of one-third solar.  The result is shown in
Figure~\ref{fig:lum_temp} along with the result for a self-similar
model (this is actually flatter than $L_X \sim T^2$ due to the
increased importance of metal lines at low temperatures).  In the same
figure, we also show the $L_X-T$ relation which would result from
keeping a self-similar shape for the gas profile, but reducing the gas
fraction as specified by eq.~(\ref{eq:fstar}).

The agreement is good, except at very low temperatures, where the
observations fall below the theoretical curve.  The surface brightness
of these poor groups is very low, and so a significant fraction of
their luminosity could be lost.  Helsdon \& Ponman (2000) estimate
that for their lowest temperature cluster group (below 1 keV), the
total flux is underestimated by about 40\% relative to the higher
temperature groups.  Also, there is some evidence that the metallicity
of small groups may be significantly less than that of larger groups
\citep{dav99}.   These two effects would reduce the predicted luminosity
by a factor of 2-3.  

There is another line of evidence which has been taken as strong
evidence of preheating: Ponman, Cannon \& Navarro (1999) show that the
central entropy (at $r=0.1 r_{200}$) in clusters and groups does not
scale in a self-similar fashion.  Their data are reproduced in
Figure~\ref{fig:entropy}, along with the curve predicted from the
model described in this paper.  The model matches the observed
data.  Also shown is the self-similar relation (with constant
$f_{gas}$).

\vspace{\baselineskip}
\epsfxsize=3.5in
\centerline{\epsfbox{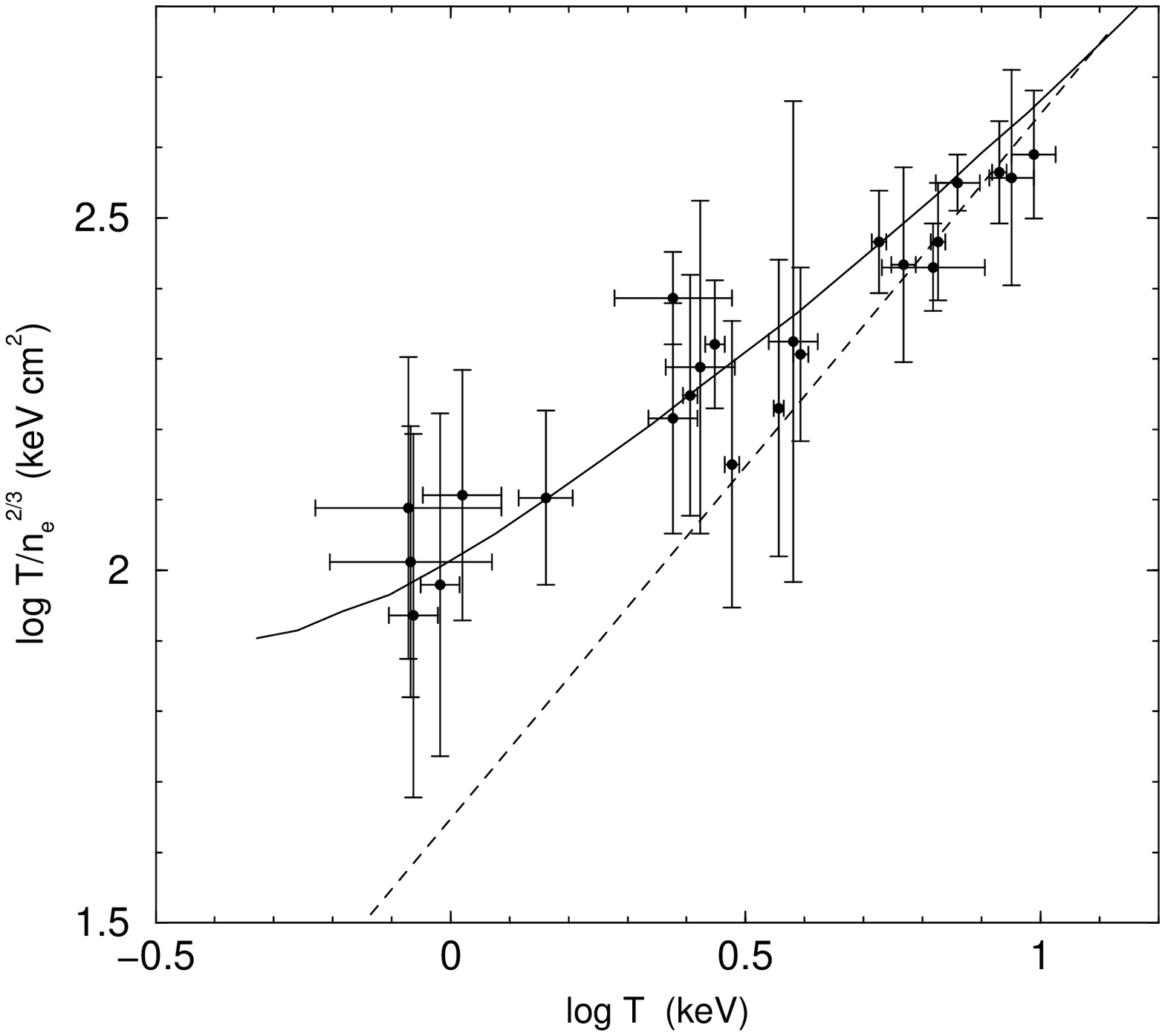}}
\figcaption{
The entropy of the gas at $r=0.1 r_{200}$ for a collection
of clusters and groups (\cite{pon99}).  The solid line shows the
prediction for the model described in this paper, while the dashed
line is the self-similar relation.
\label{fig:entropy}}
\vspace{\baselineskip}

There is a range of other observations against which this model could
be tested; we restrict ourselves to two others.  There is some
evidence that the profiles of groups is flatter than that of clusters.
For example, if the X-ray surface brightness profile is fit with a
beta model: $S(R) = S_0 (1+(R/R_c)^2)^{-3\beta + 1/2}$, then the outer
slope $\beta$ is $\sim 0.7$ for large clusters, ranging down to $\beta
\sim 0.4$ for groups \citep{moh99,hel00}.  In this
expression, $R_c$ is the projected cluster core radius.  As might be
imagined from Figure~\ref{fig:profiles}, our model also shows this
trend.  After integrating along lines of sight, the resulting surface
brightness profile is well fit by $\beta = 0.75$ for a 10.2 keV
cluster and $\beta=0.5$ for a 1.2 keV group.

The last check we make is to examine the evolution of the $L_X-T$
relation with redshift.  Observations \citep{mus97,schind99,fai00}
show that there is very little change in this relation to $z \sim
0.5$, although the amount of data are still limited.  Unfortunately,
no high redshift equivalent of Figure~\ref{fig:mratio} exists;
however, if we assume that the ratios do not change appreciably then
this models also predicts little evolution to modest redshifts.
Indeed, most models which correctly predict
$L_X \sim T^3$ will reproduce this lack of evolution.  The reason is
simple: for a fixed mass, the luminosity scales roughly as $(1+z)^3$ as
long as the profile doesn't change very much when expressed as a
function of $r/r_{200}$.  Also for a fixed mass, the virial temperature scales
as $(1+z)$ and so modifying $z$ moves a cluster parallel to the $L_X
\sim T^3$ relation.


\section{Discussion}
\label{sec:discussion}

In this letter, we have described a simple model of cluster formation
which reproduces the self-similar breaking observations without
recourse to non-gravitational heating.  There are two key assumptions
in this model.  The first is that galaxy formation was more efficient
in groups than in clusters; as discussed in section~\ref{sec:model}
there is some empirical evidence for this.  Certainly the
morphology-density relation shows that galaxies are sensitive to their
environment.  From a theoretical standpoint, this could arising from
biasing \citep{dav92} or from cooling and shocking of gas.

The second important assumption is that the lowest entropy gas is
converted into galaxies, while the high-entropy gas retains the same
entropy it would have had without galaxy formation.  Clearly this is
an idealization: in practice the rest of the gas will suffer some
radiative losses (and if it cools substantially, this will invalidate
the model assumed here).  However, the approximation is
self-consistent in that the remaining gas has a cooling time
comparable to or longer than the Hubble time.  It is also true that
cooling of hot gas tends to occur catastrophically (e.g. Thoul \&
Weinberg 1995).  That is, it remains hot with little cooling until it
passes through a cooling front, where the density suddenly increases
by orders of magnitude.  The gas at large radius moves towards the
center without changing it's entropy.  
It seems clear that numerical simulations will
be required to test these arguments, although it will be
computationally challenging to do so.  It is possible that some
of the effects described in this paper may have already been seen in
simulations \citep{pea00}.



\acknowledgements
                                                                       
We acknowledge useful discussions with Mark Voit, Joe Mohr and the
referee, Richard Bower.  Support for this work was provided by NASA
through Hubble Fellowship grant HF-01104.01-98A from the Space
Telescope Science Institute, which is operated under NASA contract
NAS6-26555.

\end{document}